\documentclass[aip,amsmath,amsfonts,amssymb,reprint]{revtex4-1}
\usepackage{graphicx}
\usepackage{epsf}
\usepackage{braket}
\usepackage{epstopdf}
\usepackage{setspace}
\usepackage{array}
\usepackage{float}
\usepackage{hyperref}
\usepackage{color}
\usepackage{subfigure}

\begin{document}
	
	\title{\textcolor{black}{Engineering two-dimensional kagome topological insulator from porous graphene 
	}}
	
	\author{Shashikant Kumar}
	\affiliation{Department of Physics,
		Indian Institute of Technology Patna, Bihta, Bihar, 801106, India}
	\author{Gulshan Kumar}
	\affiliation{Department of Physics,
		Indian Institute of Technology Patna, Bihta, Bihar, 801106, India}
	\author{Ajay Kumar}
	\affiliation{Department of Physics,
		Indian Institute of Technology Patna, Bihta, Bihar, 801106, India}
	\author{Prakash Parida}\email{pparida@iitp.ac.in}
	\affiliation{Department of Physics,
		Indian Institute of Technology Patna, Bihta, Bihar, 801106, India}
	\begin{abstract}
		Our study sets forth a carbon based two-dimensional (2D) kagome topological insulator without containing any metal atoms, that aligns the Fermi level with the Dirac
		point without the need for doping, overcoming a significant bottleneck issue observed in 2D
		metal-organic frameworks (MOFs)-based kagome structures. Our 2D kagome structure formed by creating patterned nano pores in the graphene sheet, nomenclatured as porous graphene-based kagome lattice (PGKL), is inspired by the recent bottom-up synthesis of similar structures.
		Because of absence of mirror symmetry in our porous graphene, by considering only first nearest neighbour intrinsic spin-orbit coupling (ISOC) within the tight-binding model unlike mostly used next nearest neighbour ISOC in the Kane-Mele model for graphene, PGKL exhibits distinctive band structures with Dirac bands amidst flat bands, allowing for the realization of topological states near the Fermi level. Delving into Berry curvature and Chern numbers provides a comprehensive understanding of the topological insulating properties of PGKL, offering valuable insights into 2D topological insulators. Analysis of the 1-D ribbon structure underscores the emergence of topological edge states.

	\end{abstract}
	
	\maketitle
	
	\textcolor{black}{Both graphene and kagome structures have Dirac bands, but the kagome structure uniquely includes an additional flat band\cite{jiang2019topological}. This flat band highlights the kagome lattice topological intricacies, which contribute to both topological features near the flat band and the existence of Dirac bands\cite{kim2020emergent}.}
	
		\textcolor{black}{To date, the synthesis of the kagome structure has primarily been limited to its 3D form, with examples including $Fe_3Sn_2$\cite{ye2018massive}, $CoSn$\cite{kang2020topological}, and $FeSn$\cite{kang2020dirac}, all realized as topological insulators (TIs). However, the realization of a 2D kagome lattice remains challenging, and so far, only 2D metal-organic frameworks (MOFs) based on kagome structures have been explored as TIs\cite{takenaka2021strongly, jiang2021exotic, baidya2019chern, kumar2021manifestation, guterding2016prospect, kumar2018two, zhang2019two, jiang2019dichotomy}. In most MOF-based kagome structures, the Fermi level does not align closely with the Dirac point or the flat band, making doping necessary to bring the Fermi level near these points\cite{wang2013prediction, deng20192d, gao2020design, wang2018kane, deng2021designing, yamada2016first}.}
	
\begin{figure*} 
		\centering \includegraphics[width=2.0\columnwidth]{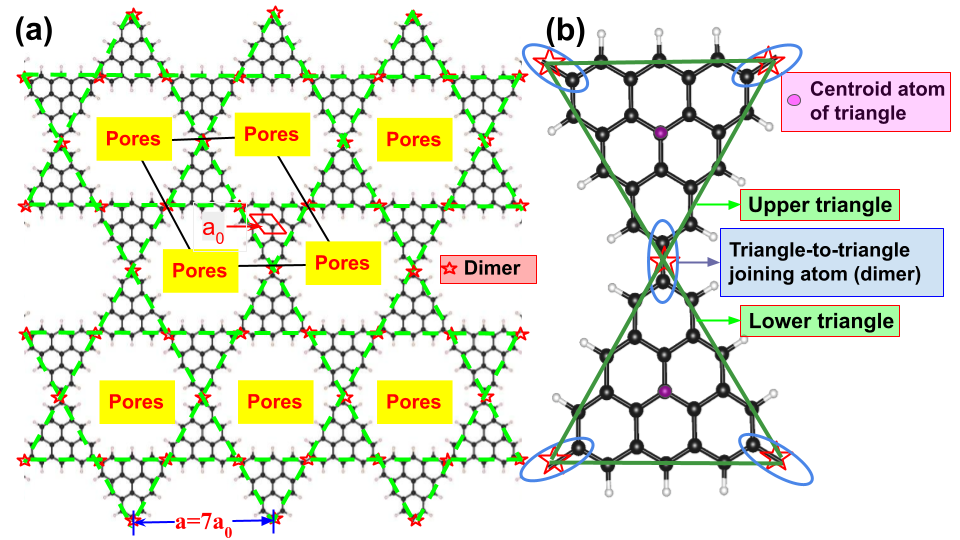}
		\caption{
			\label{fig:main_structure} 
			\textcolor{black}{{(a)} Schematic representation of the PGKL structure, with the green dashed line denoting the kagome lattice framework marked by red stars. The rhombus outlined by black border lines represents the unit cell of the structure, with a lattice constant of \( a = 7a_0 \). The red line rhombus shows the unit cell of pristine graphene, indicated by a blue arrow, where \( a_0 \) is the lattice constant of the pristine graphene unit cell, given as \( a_0 = 2.46 \, \text{\AA} \).
			{(b)} Schematic representation of atoms within the unit cell.
				}}
			\end{figure*}

	In this letter, we engineer a two-dimensional kagome structure based on purely carbon atoms (and without any metal atom), wherein the Fermi level aligns with the Dirac point without the need for any form of doping. The kagome structure is engineered by making patterned pores in graphene, resulting in a distinct configuration depicted in Fig. \ref{fig:main_structure}, referred to as the PGKL. If we carve out hexagonal pores in graphene, the remaining atoms configure into kagome structures in which carbon atoms are arranged in a dumb-bell shape in the unit cell. Fig. \ref{fig:main_structure} illustrates the distinctive kagome lattice outlined by the green dashed line. The upper and lower triangles of the dumb-bell are connected through C-C dimer represented by blue circles in the Fig. \ref{fig:main_structure}(b). These dimmers (referred as red \textcolor{red}{$\star$} mark) create a flawless kagome lattice, as highlighted by the green dashed line.
	The centroid atom of the triangle (CT) is marked in magenta colour.

	\textcolor{black}{Several nanopores have been engineered in graphene structures experimentally\cite{eroms2009weak, giesbers2012charge, sandner2015ballistic, bai2010graphene, kim2010fabrication, kim2012electronic, liu2011nanosphere, paul2012graphene, wang2013cvd, esfandiar2013dna, zhu2014plasmon, liang2010formation, zribi2016large, schmidt2018structurally} and have also been explored theoretically\cite{pedersen2008graphene, furst2009density}. Schmidt et al. highlighted helium ion beam milling as a promising method for achieving smaller pore sizes with shorter inter-defect distances\cite{schmidt2018structurally}. Their creation of pores measuring 3 to 4 nm with a pitch of nearly 10 nm demonstrated the efficacy of this approach. Recently, Jones et al. experimentally demonstrated the controlled periodic patterning of nanoscale pores in graphene\cite{jones2022nanoscale}. In their density functional theory (DFT) study, they considered a pore size of 0.7 nm. We chose a pore size of 1.5 nm in our PGKL structure to manage computational costs, which aligns with recent findings. Nevertheless, as the size and pitch of the pores increase, our main results and conclusions remain unchanged.}
		
	\textcolor{black}{In graphene sheets, the introduction of a hole or pore, commonly referred to as an anti-dot, typically results in the formation of a non-zero gap in their electronic structure \cite{pedersen2008graphene, pedersen2008optical, yuan2013electronic,blankenburg2010porous,li2010two,brunetto2012nonzero,ding2011electronic,vanevic2009character}. Addressing this issue requires the development of a different type of pore in graphene characterized by the absence of mirror symmetry while retaining electronic structures with a zero band gap. However, the realization of a zero band gap, accompanied by the emergence of Dirac and flat bands near the Fermi level, is attainable exclusively through the strategic introduction of periodic hexagonal pores within the graphene lattice. Interestingly, our designed porous structure resembles a kagome-type arrangement with no mirror symmetry. During our investigation into its topological behavior, we incorporated only first-nearest-neighbor (NN) intrinsic spin-orbit coupling (SOC).}
		
	\textcolor{black}{Moreover, researchers studying graphene have primarily focused on employing second-nearest-neighbor Kane-Mele type intrinsic SOC (ISOC) \cite{kane2005quantum, kane2005z, zhang2015prediction}. This preference stems from graphene's ability to maintain mirror symmetry, which diminishes the influence of NN-ISOC. Our implementation of NN-ISOC in PGKL is another crucial aspect of our work.}
		
	\textcolor{black}{Notably, our porous graphene design distinguishes itself from other porous variants. Our structure is supported by recent work in synthesizing very similar organic-based structures using a bottom-up approach, as demonstrated by Galeotti et al. \cite{galeotti2020synthesis} and Steiner et al. \cite{steiner2017hierarchical}.}
		
		\begin{figure*}[htbp]
			\centering
			\subfigure{\includegraphics[width=0.95\columnwidth]{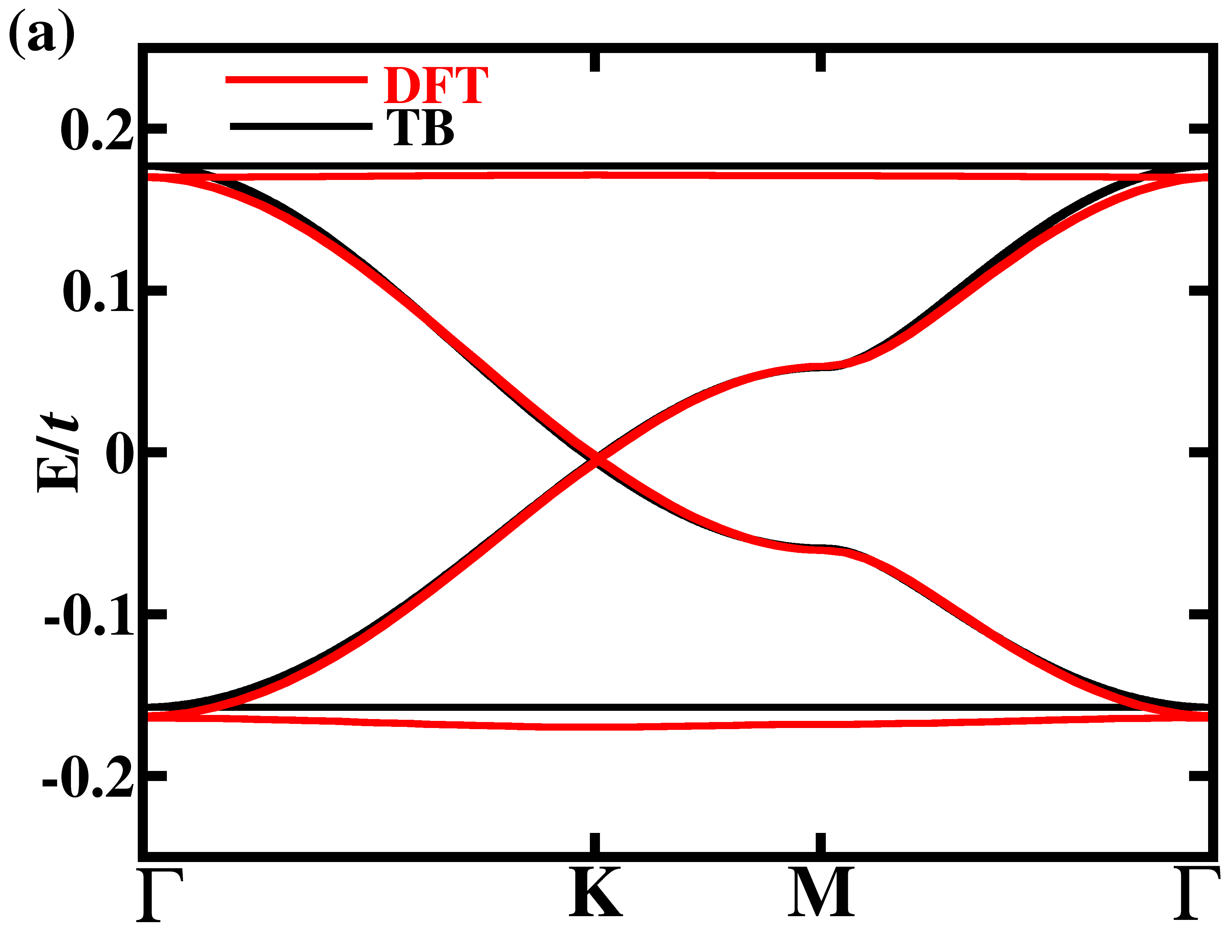}\label{fig:sub1}}
			\hspace{0.05\columnwidth} 
			\subfigure{\includegraphics[width=0.95\columnwidth]{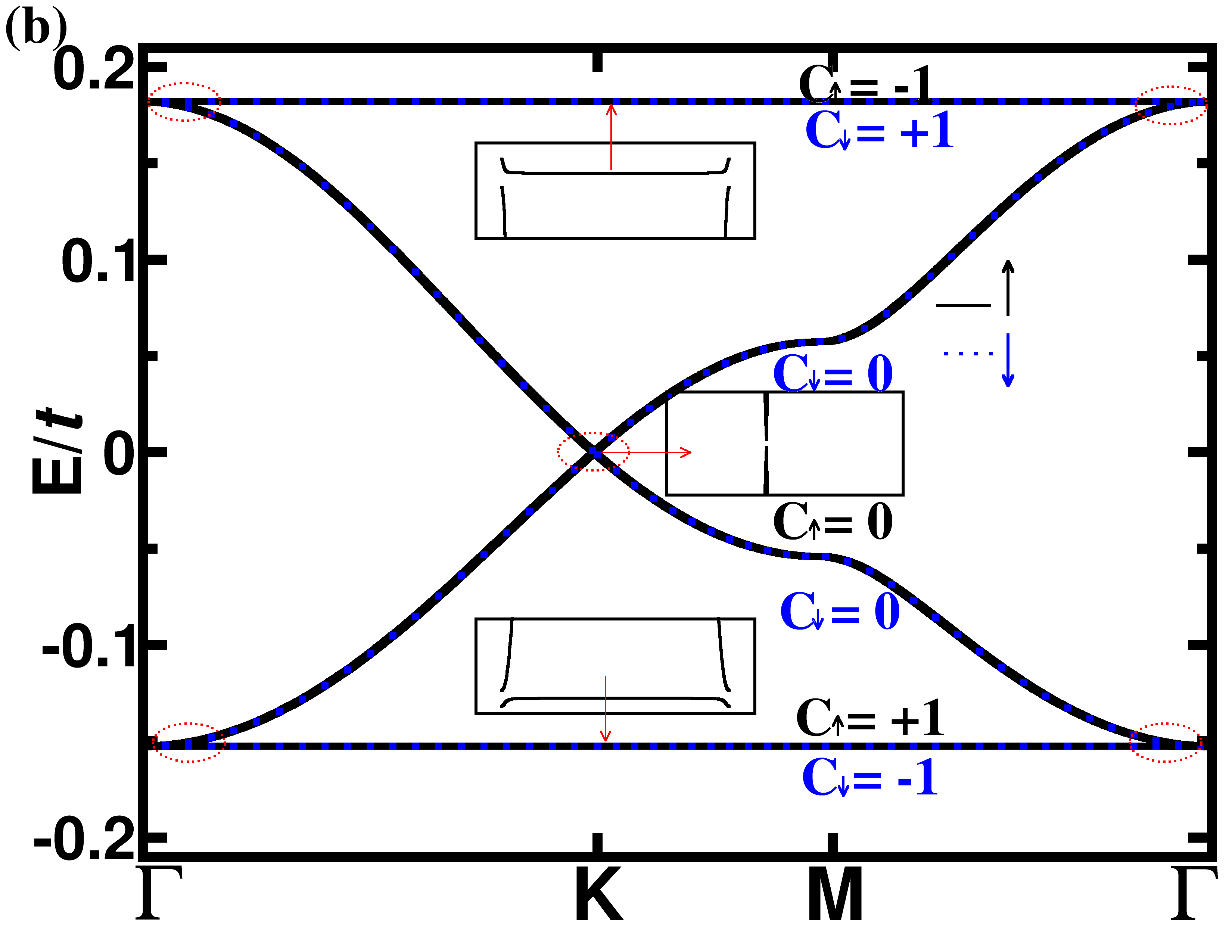}\label{fig:sub2}}
			\caption{\textcolor{black}{It represents the electronic spectra of PGKL along the high-symmetry points within the Brillouin zone. In (a), the plots depict results obtained through TB and DFT calculations without ISOC, both evaluated near the Fermi level. The DFT band structure is highlighted in red, while the TB band structure is presented in black. In (b), the representation focuses on the band structures of TB with ISOC. The parameter C$_{\uparrow/\downarrow}$ denotes the Chern number of each band. The Fermi level is scaled to zero.}}
			\label{fig:2d_PGKL_H}
		\end{figure*}
		The electronic configuration of PGKL structure shows topologically rich properties when we have studied using tight binding (TB) model (including  NN-ISOC) and density functional theory (DFT). The examination of ISOC effects reveals a non-zero spin in 2D structure, emphasizing the topological nature of the system. Furthermore, when exploring the ribbon structure of 2D configurations, we perceive the manifestation of topological edge modes. Doping the central atom with boron or nitrogen (B/N) in PGKL structure yields distinct outcomes. In 1-D ribbons, topological conducting edge modes are shown near the Dirac and $\Gamma$-points.
		
			\textcolor{black}{We have used TB model including ISOC for all structures and the corresponding Hamiltonian is given as\cite{hung2016disorder,jiang2019topological,kim2020emergent,kane2005z},}
		\begin{align}
			H=H_0 +H_{ISOC}
		\end{align}
		\begin{align}
			H_0=\sum_{i,\sigma=\uparrow,\downarrow}\varepsilon_{i}\left(c_{i\sigma}^\dagger c_{i\sigma}\right)+\sum_{<i,j>,\sigma=\uparrow,\downarrow}\left(t_{ij}c_{i\sigma}^\dagger c_{j\sigma}+h.c.\right)
		\end{align}
		\textcolor{black}{\begin{align}
			H_{ISOC}=i{\lambda_{I}\sum_{\left<i,j\right>,\sigma}\sigma c_{i\sigma}^{\dagger}\nu_{ij} c_{j\sigma}}+h.c.
		\end{align}	}

		In the provided Hamiltonian, where $H_0$ corresponds to normal TB Hamiltonian and $H_{ISOC}$ corresponds to ISOC. $c_{i\sigma}^\dagger$ represents the creation operators, and $c_{i\sigma}$ represents the annihilation operators for electrons on sites $i$. 	\textcolor{black}{Here, $\sigma$ denotes the spin states, which can be either spin up ($\uparrow$) or spin down ($\downarrow$).} The parameter $t_{ij}~=t$, represents the NN hopping amplitude from site $j$ to $i$, while $\epsilon_i$ represents the onsite electron energy at site $i$. Additionally, $\lambda_{I}$ corresponds to the strength of the NN-ISOC. The variable $\nu_{ij}$ takes value of -1 or 1, depending on whether the electron traverses the lattice in a clockwise or anticlockwise direction.
		\textcolor{black}{In equation $3$, \(\sigma = \pm 1\) indicates the spin projection along the z-axis. For spin up, the ISOC term is given by 
	\(
			H_{\text{ISOC}, \uparrow} = i\lambda_{\text{I}} \sum_{i,j} c_{i\uparrow}^\dagger \nu_{ij} c_{j\uparrow},
			\)			
			and for spin down, it is given by 			
			\(
			H_{\text{ISOC}, \downarrow} = -i\lambda_{\text{I}} \sum_{i,j} c_{i\downarrow}^\dagger \nu_{ij} c_{j\downarrow}\)\cite{hung2016disorder}.
			}
	
		At first, we have employed the TB Hamiltonian without ISOC ($H=H_0$) to characterize the electronic structure of the PGKL. By fitting the parameters through DFT [for computational details, see detailed discussion in section 1 of the supplementary material SM], we set the NN hopping strength, $t$ = -3.0 eV\cite{pereira2009tight,  stauber2008optical} and the onsite energy for each bulk atom to zero ($\epsilon_{bulk}$ = 0 eV) for all structures. Conversely, in PGKL, onsite energy for each edge atom is set as $\epsilon_{edge}$ = -3.55 eV. The resulting TB band structures, in conjunction with DFT, are plotted around the Fermi level, as depicted in Fig. \ref{fig:2d_PGKL_H}(a). For entire band structure, see Fig. S1 in SM. A notable observation is quite well matches of the TB and the DFT bands around the Fermi level. It is important to mention that, for consistency, the Fermi level is scaled to zero across all band structures. From Fig. \ref{fig:2d_PGKL_H}(a), a notable unfolds, featuring two Dirac bands flanked by two flat bands, with the flat bands touching the Dirac bands near the $\Gamma$ point. The origin of the Dirac band is deeply entwined with an underlying honeycomb structure, while the coexistence of the Dirac band with a flat band is rooted in the structural characteristics of the kagome structure. \textcolor{black}{The Dirac bands in PGKL become flatter compared to graphene, leading to a decrease in Fermi velocity. A detailed discussion with varying pore sizes is given in SM [see section 2 and Fig. S3 in the SM]. Gregersen et al. also reported that, the Fermi velocity in graphene reduces as the antidot size increases\cite{gregersen2015graphene}.}
			
	\textcolor{black}{In our pursuit to unravel the intricacies of these bands, we have adopted a systematic approach by setting first the onsite energy of both bulk and edge atoms to the same value and zero, $\epsilon_{\text{bulk}}  =  \epsilon_{\text{edge}} = 0 $. Interestingly, in close proximity to the Fermi level, all four bands exhibit a remarkable transformation, becoming flat and seamlessly superimposed [see Fig. S2 in SM]. Significantly, at the Fermi level, all four flat bands are derived from the substantial contributions of the edge atoms. Contrasting both above and below the Fermi level, there emerge two Dirac bands accompanied by a single flat band. Notably, the flat band touches one of the Dirac band near the $\Gamma$ point, as illustrated in Fig. S2 [see in SM]. } 
		\begin{figure*}[htpb]
		\centering \includegraphics[width=2.0\columnwidth]{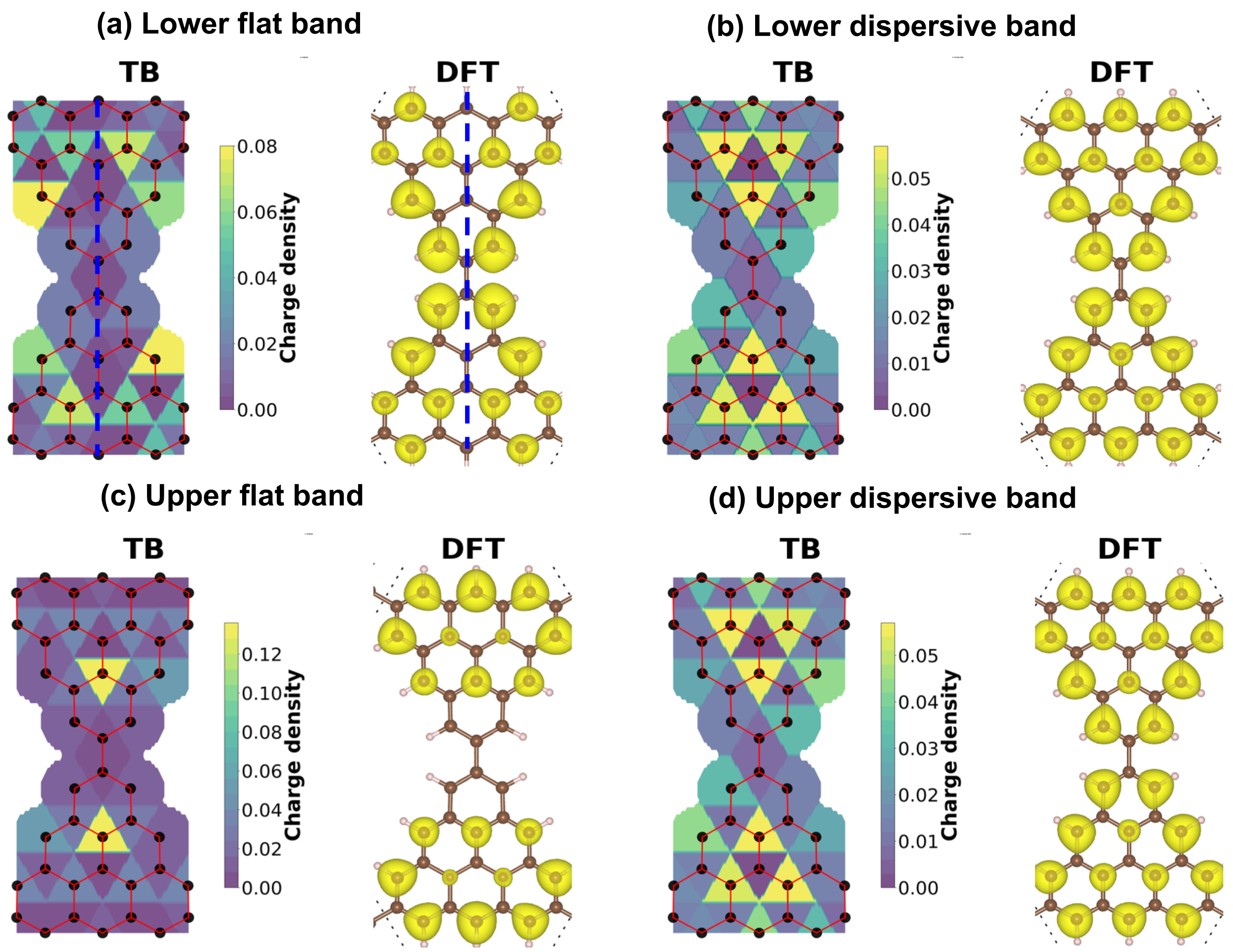}
		\caption{
			\label{fig:main_dos_4} 
			\textcolor{black}{The left panels (a) and (c) display the charge density of the lower and upper flat bands, respectively, near the Fermi level for both TB and DFT. In (a), the blue dotted line indicates the nodal line with respect to which the state is localized to the left and right of it. In (c), the flat band is clearly localised in the upper and lower portions of the upper and the lower triangles, respectively. The right panels (b) and (d) show representations of the charge density for the lower and upper dispersive bands, respectively, near the Fermi level for both TB and DFT. It is clearly visible that, both dispersive bands are delocalised over whole structure with dominant contributions from the NNCT atoms.}
		}
		
	\end{figure*}
	
	\textcolor{black}{This particular configuration is reflective of the characteristic patterns exhibited by the kagome lattice\cite{kim2020emergent}. Notably, these bands exhibit a heightened influence from the C-C dimmers (represented by \textcolor{red}{$\star$} in Fig. \ref{fig:main_structure}), a finding substantiated by the insightful charge density plot showcased in Fig. S2 in SM. These dimmers, in their structural arrangement, impeccably fashioned the kagome lattice structures, exemplified in Fig. \ref{fig:main_structure}. Our proposed structure stands as an exemplary 2D carbon-based kagome structure without any metal atom.
	Subsequent to this, we have proceeded to fine-tune the onsite energy of the edge atom, accounting for any disparity between the bulk and edge atoms. Now, we set  $\epsilon_{bulk} = 0$ eV and $\epsilon_{edge} = -10^{-4}$ eV. By doing so, we introduce an infinitesimally small disparity in the onsite energy between bulk and edge atoms ($\delta_{\epsilon}=|{\epsilon_{bulk}-\epsilon_{edge}|=10^{-4} eV}$). Remarkably, the four initial flat bands underwent a trans-formative evolution, transitioning into two flat bands and two Dirac bands. This intriguing is visually represented in Fig. S4(b), focusing on the zoomed section near the Fermi-level. In these bands, the primary contribution emanates from the edge atoms [see Fig. S4(e) in SM].} 
	
	\textcolor{black}{Additionally, apart from the edge atoms, minor contributions are observed from the three nearest neighbor atoms of the centroid atom of the triangle (NNCT). When this disparity is too large ($\delta_{\epsilon}=10^{4}$ eV, $\epsilon_{edge}=-10^4$ eV) it unveils a noteworthy observation that the same four bands (two Dirac bands and two flat bands) linger around Fermi level like the case of $\delta_{\epsilon}=10^{-4} eV$. This is a surprising result as it is expected that, these four bands should be obligated and pushed deep into the valence band for a huge negative value of onsite energy of edge atoms. However, these four bands persistently remain near the Fermi level without descending to lower energy levels. This is because, the primary contribution of these bands comes from the NNCT atoms [see Fig. S4(f)], with no discernible contribution from edge atoms. All the edge atoms form corresponding bands located well below the Fermi level (near E = $-10^4$ eV, refer to Fig. S4(c) in SM). This intriguing behavior: non-removal of these four zero-energy bands, suggests inherent structural symmetry embedded within this specific configuration.}
				
		In Fig. \ref{fig:2d_PGKL_H}(a), we now set the onsite energy of the edge atom to a reasonable value, $\epsilon_{edge}$ = -3.55 eV, while maintaining that of the bulk atoms at zero. \textcolor{black}{Our exploration of these bands involves a comprehensive analysis through charge density plots obtained from TB and DFT. Notably, NNCT atoms predominantly contribute to the low energy Dirac bands, which are shown in Fig. \ref{fig:main_dos_4}.}
		These three NNCT atoms marked as a ring [see Fig. S5 in SM], interconnect to form a flawless and periodic honeycomb structure. Thus, the emergence of Dirac bands is attributed to the inherent honeycomb arrangement of these neighboring NNCT atoms. Two additional flat bands, one in the conduction and the other in the valence band, result from destructive interference among different hopping directions at NNCT atoms. In essence, these bands exhibit characteristics reminiscent of kagome-type flat band structures. 
		
		\begin{figure*}
			\centering \includegraphics[width=2.0\columnwidth]{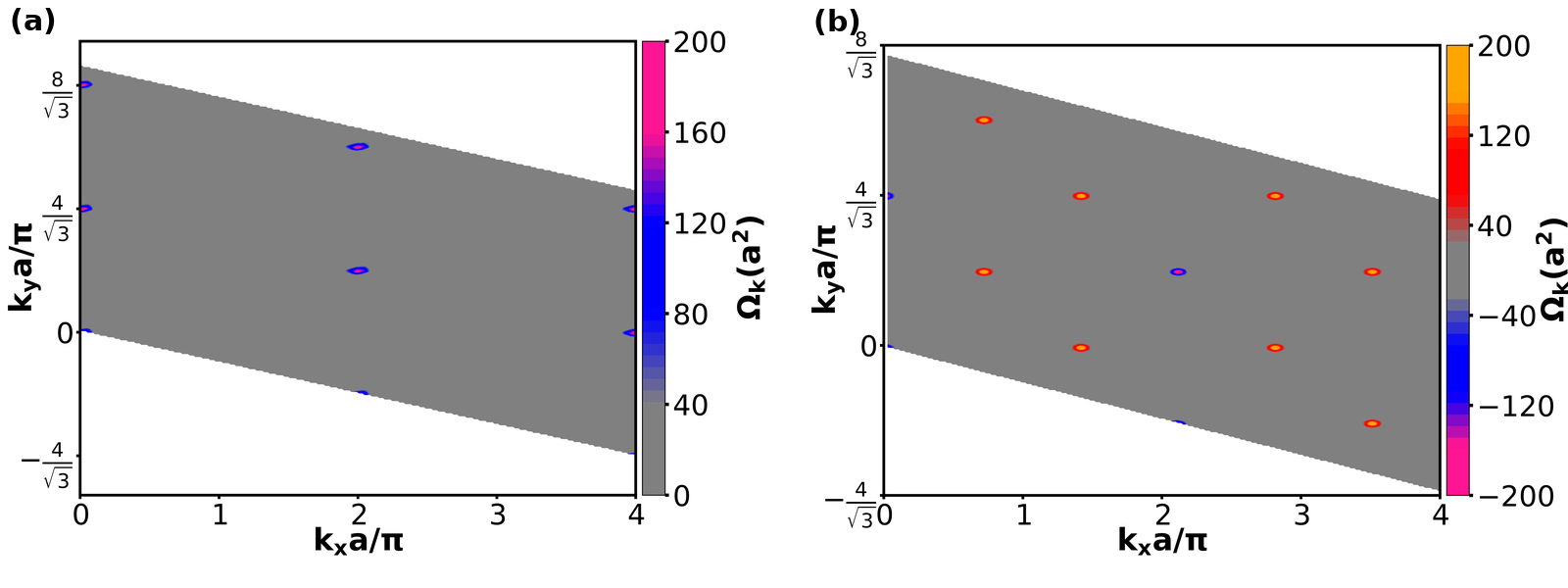}
			\caption{
				\textcolor{black}{The Berry curvature plot of the lower flat band near the Fermi level in the extended Brillouin zone (BZ), showing peaks around the $\Gamma$-points. (b) The Berry curvature plot of the lower Dirac band near the Fermi level, showing peaks around the $\Gamma$-point as well as around the $K$-points. The Berry curvature forms a hexagonal lattice, with peaks around the $K$ and $K'$ points. }
				\label{fig:berry}
			}
		\end{figure*}	
		
		To explore the topological properties, we include NN-ISOC in Hamiltonian as $H~=~H_0+H_{ISOC}$. We set the ISOC strength for NN as $\lambda_{I}$ = -1.3 meV\cite{zhang2015prediction}.
		Incorporating ISOC leads to the emergence of energy gap of 0.26 meV at the $\Gamma$ point ( $\Gamma$-gap) and 0.45 meV at the Dirac point (K-gap), effectively transitioning the system from being a semi-metal to an insulator\cite{kane2005quantum,jiang2019topological,kim2020emergent}, depicted in Fig. \ref{fig:2d_PGKL_H}(b), \textcolor{black}{where the time-reversal symmetry is preserved and the topology is characterized by the Chern number.} 	\textcolor{black}{In the absence of the Rashba potential, the Kane-Mele (KM) model maintains the $z$ projection of the spin\cite{kane2005quantum,kane2005quantum}. For any nonzero value of $\lambda_{I}$, each spin projection behaves like a quantum spin Hall insulator with opposite spin Chern number, $C_{\uparrow/\downarrow} = \pm1$, for the respective spin projections. Thus, the KM model exhibits a $Z_2$ topologically nontrivial phase when $\lambda_{I}$ is nonzero\cite{hung2016disorder,wang2016topological}.}
		
	 Nevertheless, when we adjust the strength of the ISOC from milli-electron volts (meV) to micro-electron volts ($\mu$eV), we observe that our results remain consistent. The sole discernible change is the size of the band gap; nonetheless, { our conclusions regarding the band structure remain unchanged. One way to achieve measurable band gap is by adding SOC  effects through combining graphene with semiconductor transition metal dichalcogenide substrates like $WSe_2$\cite{xiao2012coupled,zhu2011giant}. These substrates have special qualities, such as a strong SOC effect (about 500 meV in the valence band) and a broad band gap that matches well with graphene's Fermi level\cite{sun2023determining}.} The Chern number of the n$^{th}$ band with spin is defined in the first Brillouin zone (BZ) \cite{thouless1982quantized,yao2004first},
	\textcolor{black}{	\begin{eqnarray}
		{ C}_{n,\sigma} =  \frac{1}{2\pi}\int_{1^{st}BZ} \Omega_{n,\sigma}({\bf k}) d^2{\bf k} \end{eqnarray}}
	
	\textcolor{black}{Where  $\Omega_{n,\sigma}({\bf k})$ is the Berry curvature, which is defined as,}
	\textcolor{black}{\begin{align} 
		\Omega_{n,\sigma}({\bf k}) = -2\, \text{Im} \sum_{\lambda \neq n} 
		\frac{
			\langle \lambda,{\sigma} | \nabla_{\bf k} H | n,{\sigma} \rangle
			\langle n,{\sigma} | \nabla_{\bf k} H | \lambda,{\sigma} \rangle
		}{
			\left(E_{\lambda,{\sigma}} - E_{n,{\sigma}}\right)^2
		}
	\end{align}}
\textcolor{black}{Here, n is the band index with spin, while $\ket{n,{\sigma}}$ and $E_{n,{\sigma}}$ are the eigenstate
	and eigenvalue of the $n^{th}-$band with spin, respectively.} $\nabla_{\bf k}$ is the gradient vector, H is the Hamiltonian matrix. 
		
		A notable theoretical prediction by Wang et al.\cite{wang2013organic} introduced the concept of the first organic topological insulator (TI) within a honeycomb lattice structure. In this scenario, the MOFs of $Bi_2(C_6H_4)_3$ and $Pb_2(C_6H_4)_3$ serve as the platform for the Dirac bands with intriguing electronic properties\cite{wang2013prediction,wang2013organic}. The theoretical implementation of this TB model was first achieved within MOFs, specifically starting with $In_2(C_6H_4)_3$\cite{liu2013flat} and subsequently extended to $Tl_2(C_6H_4)_3$\cite{10.1063/1.5017956}. In our Fig. \ref{fig:2d_PGKL_H}(b), all three gaps hold topologically significant properties. \textcolor{black}{This is discerned through the distinctive band-specific spin Chern numbers assigned to each gap near the Fermi level, in a descending order for $C_{n,\uparrow}: 1, 0, 0, ~and -1$, or ascending order for $C_{n,\downarrow}: -1,0,0,~and +1$. Remarkably, the introduction of ISOC results in a subtle dispersion of the flat bands. \textcolor{black}{The spin Chern number ($C_{n,s}$), defined as $\frac{C_{n,\uparrow} - C_{n,\downarrow}}{2}$, where $C_{n,\uparrow}$ and $C_{n,\downarrow}$ represent the spin Chern numbers for the $\uparrow$ and $\downarrow$ bands respectively\cite{zhang2019topologically}.}}
			\begin{figure*}
			\centering \includegraphics[width=2.0\columnwidth]{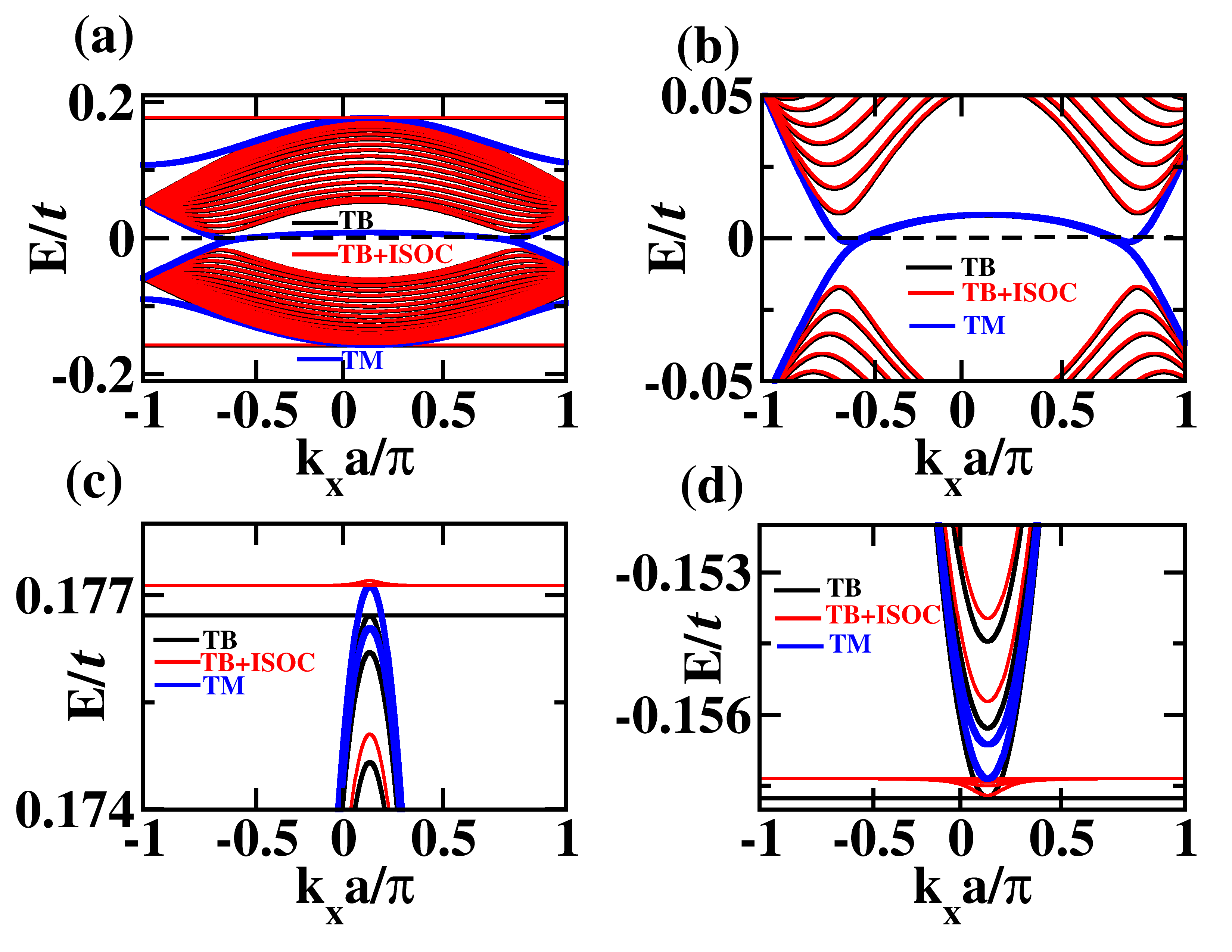}
			\caption{The band structure of the 1-D zigzag-like ribbon of PGKL is depicted, with panels (a), (b), (c), and (d) representing the regions near, at, above, and below the Fermi level, respectively. The black and red lines illustrate the band structures without and with ISOC, respectively. The blue line corresponds to the topological modes (TM), while the black dotted line indicates the Fermi level. The Fermi level is scaled to zero.
				\label{fig:ribbon} 
			}
		\end{figure*}
	
		In Fig. \ref{fig:2d_PGKL_H}(b), it manifests that below the energy gaps exhibit non-trivial topological behavior. This fact is underscored by the plot of the Berry curvature presented in Fig. \ref{fig:berry}. The plot unequivocally showcases the localization of non-zero Berry curvature, particularly in proximity to the $\Gamma$ (flat band) and K-gap (Dirac band). 
	    The cumulative insights derived from the Berry curvature plot and the spin Chern number confirm the non-trivial topological insulating characteristics of the PGKL structure [see Fig. 4].
		Interestingly, our exploration reveals an exquisite congruence of Berry curvature between the PGKL and the pristine kagome lattice	\cite{jiang2019topological}, affirming the robustness of our findings. Our findings align closely with their presentation, highlighting the significance of the Berry curvature phenomena in organic TIs within organo-metallic lattices \cite{wang2013prediction,liu2013flat}.
		\textcolor{black}{When CT atoms are doped with B/N atoms, the resulting band structure is illustrated in Fig. S6 in SM [see detailed discussion in section 3 of the SM].}
		 
		Further, we analyze the band structure of 1-D ribbons, expecting topological conducting edge modes to appear within the K-gap and $\Gamma$-gap. We have constructed zigzag-like and armchair-like 1-D ribbons [see Fig. S8 in SM]. Among these two types of ribbons, we first study the topological properties of zigzag-like ribbon. Our study revolves around the examination of the band structure near the Fermi level. In instances where ISOC is absent, a discernible zero-gap emerges between dispersive bands, as depicted in Fig. \ref{fig:ribbon}. Upon the introduction of ISOC, this gap between dispersive bands and flat bands widens substantially. Along with, gap-less topological edge modes appear in the SOC-gap crossing the Fermi level [see the blue shaded bands in Fig. \ref{fig:ribbon}]. Thus, PGKL is a TI having a SOC gap in its bulk and conducting edge modes closing the gap. Our band structure matches with the previously predicted band structure in graphene and kagome ribbons\cite{van2020volkov,do2010graphene,kim2020emergent,jiang2019topological}. 
		
		Each gap within these ribbon structures bears topological significance, underpinned by the non-zero total spin Chern numbers below these gaps, as shown in Fig. \ref{fig:2d_PGKL_H}(b). 
		Notably, the presence of non-zero spin Chern numbers beneath these gaps ensures the topological nature of the edge modes. Conversely, if the cumulative spin Chern numbers become zero below a given gap, the corresponding edge modes revert to trivial insulators. In the ribbon configuration, doping B/N at the CT allows access to the Fermi level near the flat band, where the dispersive band and flat band are connected through topological modes, as shown in Fig. S9(c) and S9(d) [see Fig. S9 in SM]. However, for armchair like ribbon, unlike zigzag-like, no conducting edge modes appear in the K-gap. Moreover, $\Gamma$-gap is closed by topological edge modes [see Fig. S10 in SM]. It is worth noting that, zigzag-like ribbon shows superior topological properties compared to armchair-like.
		
		In this work, our study explores the PGKL, a 2D carbon-based kagome TI that uniquely aligns the Fermi level with the Dirac cone without the requirement of any doping. This structure exhibits a distinctive band structure with Dirac bands amidst flat bands, allowing for the realization of topological states near the Fermi level. Our analysis of 1-D ribbon structures showcased the emergence of topological edge states, emphasizing the resilience of the structure to external perturbations. The examination of Berry curvature and Chern numbers provide a robust understanding of the topological insulating properties of PGKL, contributing valuable insights to the field of 2D TIs. PGKL introduces an addition to 2D kagome structures, promising versatility and topological richness for possible potential applications.
		
		\section*{Supplementary Material }
		The supplementary material comprises computational details, band structures of PGKL. Additionally, detailed analyses of armchair-like and zigzag-like ribbon structures, including their respective band structures, are provided.
		
		A.K. acknowledges University Grants Commission (UGC), New Delhi, for financial support in the form of a Senior Research Fellowship (No. DEC18-512569-ACTIVE). P.P. acknowledges DST-SERB for the ECRA project (No. ECR/2017/003305).\\
		\section*{Conflict of interest}
		The authors declare no conflict of interest.
		\section*{Author Contributions}
		\textbf{Shashikant Kumar}: Conceptualization (equal); Data curation (equal); Formal analysis (equal); Investigation (equal); Methodology (equal); Writing – original draft (equal); Writing – review and editing (equal). 
		\textbf{Gulshan Kumar}: Conceptualization (supporting); Formal analysis (supporting); Visualization (equal); Writing – review and editing (supporting). 
		\textbf{Ajay Kumar}: Formal analysis (supporting); Software (supporting); Visualization (supporting); Writing – review and editing (supporting). \textbf{Prakash Parida}:   Conceptualization (equal); Supervision (equal); Validation (equal); Visualization (equal); Writing – review and editing (equal).
		
		\section*{DATA AVAILABILITY}
		The data that support the findings of this study are available from the corresponding author upon reasonable request.

		\bibliography{main.bib}
		
	\end{document}